\documentclass[aps,prl,twocolumn,groupedaddress]{revtex4-2} 

\usepackage{bm}
\usepackage{color}
\usepackage{graphicx}
\usepackage{amsmath}
\usepackage{braket}
\usepackage{changes}

\begin{document}

\title{Full monitoring of phase-space trajectories with 10\,dB-sub-Heisenberg imprecision}
\author{Jascha Zander}
\affiliation{Institut f\"ur Laserphysik und Zentrum f\"ur Optische Quantentechnologien, Universit\"at Hamburg, Luruper Chaussee 149, 22761 Hamburg, Germany}

\author{Roman Schnabel}
\email{roman.schnabel@uni-hamburg.de}
\affiliation{Institut f\"ur Laserphysik und Zentrum f\"ur Optische Quantentechnologien, Universit\"at Hamburg, Luruper Chaussee 149, 22761 Hamburg, Germany}

\date{April 12, 2021}

\begin{abstract}
The change of a quantum state can generally only be fully monitored through simultaneous measurements of two non-commuting observables $\hat X$ and $\hat Y$ spanning a phase space. A measurement device that is coupled to the thermal environment provides at a time a pair of values that have a minimal uncertainty product set by the Heisenberg uncertainty relation, which limits the precision of the monitoring. 
Here we report on an optical measurement setup that is able to monitor the time dependent change of the quantum state's displacement in phase space ($\langle \hat X (t) \rangle; \langle \hat Y (t) \rangle$) with an imprecision 10\,dB below the Heisenberg uncertainty limit. Our setup provides pairs of values ($X(t_i); Y(t_i)$) from simultaneous measurements at subsequent times $t_i$. The measurement references are not coupled to the thermal environment but are established by an entangled quantum state. 
Our achievement of a tenfold reduced quantum imprecision in monitoring arbitrary time-dependent displacements supports the potential of the quantum technology required for entanglement-enhanced metrology and sensing as well as measurement-based quantum computing.
\end{abstract}

\maketitle

\emph{Introduction} -- Quantum sensing and measurement-based quantum computing utilize quantum correlated states \cite{Einstein1935,Schroedinger1935}. The most mature technology for these applications are based on quantum states of the electro-magnetic field. Laser interferometers that are used as telescopes for gravitational-wave astronomy achieve unprecedented sensitivities based on states having a squeezed photon counting statistic \cite{LSC2011,Tse2019, Acernese2019,Schnabel2010}. It was proposed to further improve them by bi-partite Gaussian entangled states \cite{SteinlechnerS2013,Ma2017,Suedbeck2020,Yap2020}.
Optical measurement-based quantum computing based on multi-partite entangled cluster states \cite{Raussendorf2001,OBrian2007} was pushed forward recently \cite{Larsen2019}.

Measurements in the regime of Gaussian quantum statistics concern two non-commuting observables. 
In terms of dimensionless operators that are normalised to the variance of the ground state, they are often named $\hat X$ and $\hat Y$.  
They need to be measured both in order to determine the full energy of a quantised harmonic oscillator, similar to position and momentum.
The Heisenberg uncertainty relation \cite{Heisenberg1927,Kennard1927,Weyl1927,Robertson1929} is a useful reference to distinguish between semi-classical measurements \cite{Arthurs1965} and those that exploit entanglement \cite{DAriano2001}.
Experimentally achievable Gaussian entanglement have been characterised by co-variances derived from ensemble measurements in stationary settings \cite{Ou1992,Bowen2003a,SteinlechnerS2013,Eberle2013}. So far, Gaussian entanglement was not used to improve measurements of phase-space displacements that changed after a single measurement window. In such a time-dependent setting, averaging would not constitute a suitable approach for improving the signal-to-noise-ratio.  

Here, we present for the first time the monitoring of a dynamical phase-space trajectory $\alpha (t)$ through the simultaneous measurements of two non-commuting observables ($X(t_i);Y(t_i)$) at subsequent times $t_i$ with an imprecision much lower than the reference limit as given by the Heisenberg uncertainty relation. The nonclassical improvement provides the same benefit as ten-times averaging, which, however, is possible only in a stationary setting. The experimental achievements presented here are entirely based on individually sampled two-dimensional data points.  

\begin{figure}[t!!!!!]
     \vspace{1mm}
        \hspace{0mm}\includegraphics[width=8.6cm]{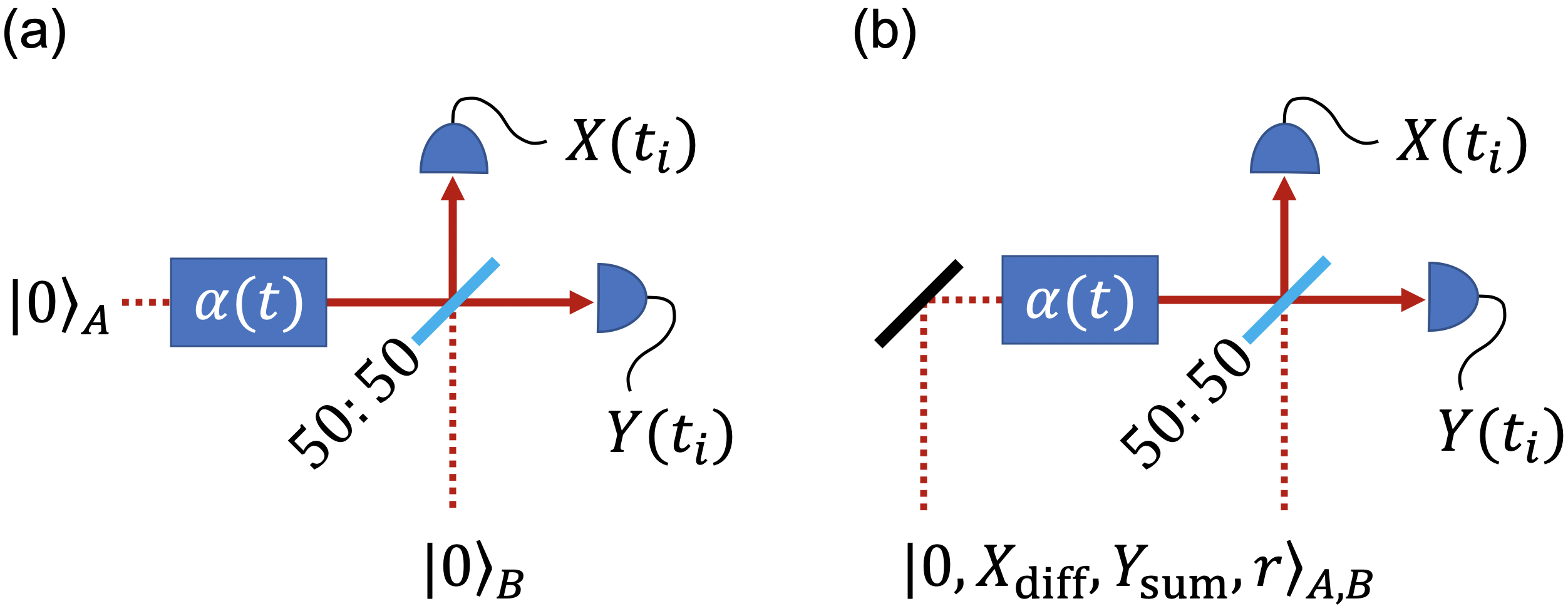}
     \vspace{-3mm}
    \caption{Monitoring of quantum state displacement $\alpha (t)$. The phase space observables are the real and imaginary parts $\alpha (t) = \langle \hat X (t) \rangle + i \langle \hat Y (t) \rangle$. (a) The simultaneous (semi-classical) monitoring of the displacement is hampered by quantum uncertainties, which can be illustrated by ground states entering the measurement device at inputs $A$ and $B$. (b) Exploiting an entangled state $|0, X_{\rm diff}, Y_{\rm sum}, r \rangle_{A,B}$ reduces the quantum noise imprecision of a single measurement pair ($X(t_i); Y(t_i)$) at time $t_i$ in principle to arbitrarily low values. Here, `$0$' refers to the average displacements at inputs A and B, $X_{\rm diff}, Y_{\rm sum}$ to the kind of quantum correlations, and $r$ to the joint strength of the quantum correlations. In practice, decoherence on the entangled state that reduces $r$ is the major problem.
}
     \vspace{0mm}
    \label{fig:1}
\end{figure}
%

\begin{figure}[t!!!!!!!!!!!!!!!!!]
     \vspace{0mm}
    \includegraphics[width=8.75cm]{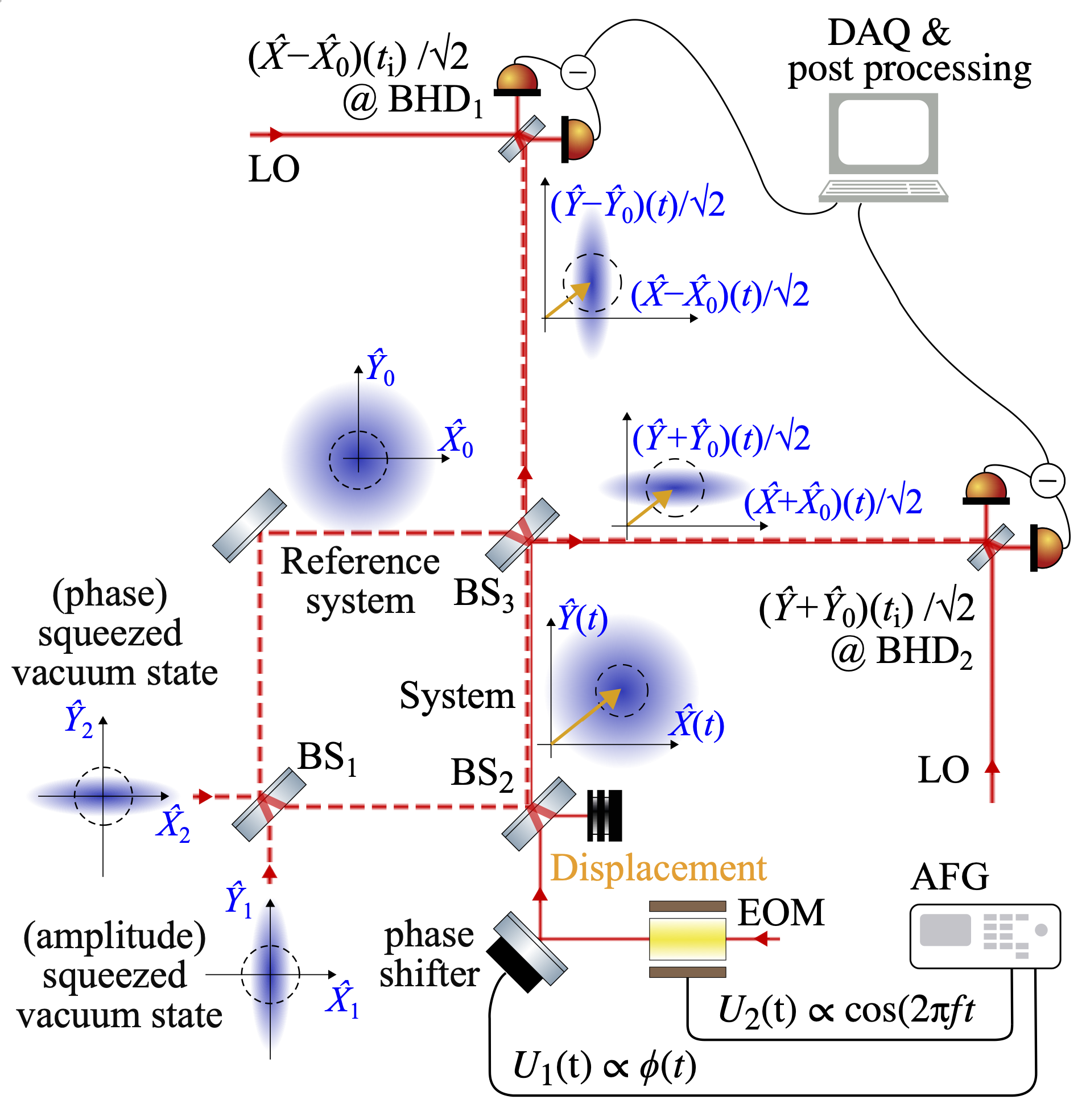}
         \vspace{-4mm}
    \caption{
Schematic of the experiment. Shown are optical paths of laser beams at wavelength 1550\,nm. The phase-space pictures show the quantum uncertainties of the laser beams' modulations at $f \!=\! 5$\,MHz at several instances in the Wigner representation. From bottom left to top right: balanced beam splitter BS\textsubscript{1} converted two squeezed vacuum states into a bipartite EPR entangled state. One part served as a quantum reference (subscript $0$). The other part was displaced by $(\langle \hat X \rangle; \!\langle \hat Y \rangle) (t)$ (illustrated by the arrow) by overlapping modulated light transmitted through BS$_{\rm 2}$. 
The two projections of the arrow were simultaneous monitored with respect to the entangled reference system by superposition at BS\textsubscript{3} and by detecting the outputs with balanced homodyne detectors. BHD\textsubscript{1} provided eigenvalues of $(\hat X \!-\! \hat X_0)(t_i) / \sqrt{2}$, while BHD\textsubscript{2} provided eigenvalues of $(\hat Y \!+\! \hat Y_0)(t_i) / \sqrt{2}$, with $\langle \hat X_0 \rangle \!=\! \langle \hat Y_0 \rangle \!=\! 0$. 
EOM: Electro-Optical Modulator, AFG: Arbitrary Function Generator, DAQ: Data AcQuisition, LO: Local Oscillator.
}
         \vspace{-2mm}
    \label{fig:2}
\end{figure}

\textit{Uncertainty relations} -- 
Our experiment uses quantities that are also used in optical communication and optical quantum computing \cite{Larsen2019}, namely phase and amplitude modulation depths carried by quasi-monochromatic laser light of optical frequency $\nu$, see the figure in the supplement.
The depth of the amplitude modulation (amplitude modulation index) in the frequency band $f \pm \Delta f$, with $\nu \!\gg\! f \!>\! \Delta f$, is quantified by the dimensionless operator $\hat{X}_{f, \Delta f}$ \cite{Schnabel2017}. This operator is also known as `amplitude quadrature amplitude'. 
The corresponding depth of phase modulations is (in the limit of weak phase modulations) quantified by the operator $\hat{Y}_{f, \Delta f}$. This operator is also known as `phase quadrature amplitude' \cite{Caves1985a}. $\hat{X}_{f, \Delta f}$ and $\hat{Y}_{f, \Delta f}$ do not commute. 
In the following we skip the indices and normalise the commutator to $[\hat{X},\hat{Y}] = 2i$, which results in the Heisenberg uncertainty relation
      \begin{equation}
      	\Delta \hat X \Delta \hat Y \ge 1 \, , 
	\label{eq:hup1}
      \end{equation}
where $\Delta$ denotes the standard deviation of the measured eigenvalues of the respective operator. $\hat{X}$  and $\hat{Y}$ span a phase-space, in which the uncertainty area is bounded from below accordingly. The lower bound in Eq.\,(\ref{eq:hup1}) refers to `ideal' measurements performed with semi-classical devices, which do not use quantum correlations. `Ideal' means that $\hat X$ is measured on a first copy and $\hat Y$ is measured on a second copy of the system. Ideal measurements are only possible in (quasi-)stationary settings.\\
If the quantities $\hat{X}$ and $\hat{Y}$ change on time scales that are not much longer than $1/\Delta f$, they need to be measured simultaneously.
In this case, splitting the system into two independently measured subsystems is required (beam splitter in Fig.\,\ref{fig:1}, and BS$_3$ in Fig.\,\ref{fig:2}). 
Furthermore, averaging is not possible unless the trajectory repeats after some time.
The splitting reduces the signal-to-noise ratio in comparison with ideal measurements. The splitting can be described as opening a new port through which vacuum uncertainty couples to the measurement (port $B$ in Fig.\,\ref{fig:1}, left), if not an entangled reference system is superimposed via this port, as shown in in Fig.\,\ref{fig:1} (b) and Fig.\,\ref{fig:2}.
In the absence of quantum correlations, simultaneous measurements at times $t_i$ need to cope with at least doubled minimal quantum uncertainties, which increases standard deviations by at least the factor $\sqrt{2}$ \cite{Arthurs1965}, yielding
      \begin{equation}
          \Delta \left( \hat X (t_i) \right) \Delta \left( \hat Y (t_i)\right) \geq 2 \, .
          \label{eq:hup2}
      \end{equation}
The above inequality represents the fundamental precision limit when two conjugate observables are measured simultaneously on a single system with respect to reference values of a semi-classical measurement device.
Note that Inequality\,(\ref{eq:hup2}) relates to a Gaussian state in the Husimi Q representation \cite{Husimi1940}, whereas Inequality\,(\ref{eq:hup1}) relates to the Wigner representation \cite{Wigner1932}.  

\emph{Measurements with respect to quantum references} -- 
Let $[\hat{X},\hat{Y}] \!=\! 2i$ describe a quantum system of interest, and $[\hat{X}_0,\hat{Y}_0] \!=\! 2i$ another quantum system.
A short calculation leads to the zero-commutator $[\hat X \!\pm\! \hat X_0, \hat Y \!\mp\! \hat Y_0] \!=\! 0$, which results in 
      \begin{equation}
      	\Delta(\hat X \pm \hat X_0) \,\Delta(\hat Y \mp \hat Y_0) \ge 0 \, .
	\label{eq:hup3}
      \end{equation}
This inequality describes the fact that $X$ and $Y$ of a system can be measured simultaneously with arbitrary precision with respect to the corresponding quantities $X_0$ and $Y_0$ of a reference system. Usually, a reference system has its own quantum uncertainty. 
If, however, an ensemble of quantum systems are available that are all entangled with a reference system at hand, the measurement of a phase-space trajectory $(\langle \hat X \rangle; \!\langle \hat Y \rangle) (t)$ with a sub-Heisenberg imprecision is possible.

\begin{figure}[]
     \vspace{0mm}
     \hspace{-2mm}
    \includegraphics[width=8.7cm]{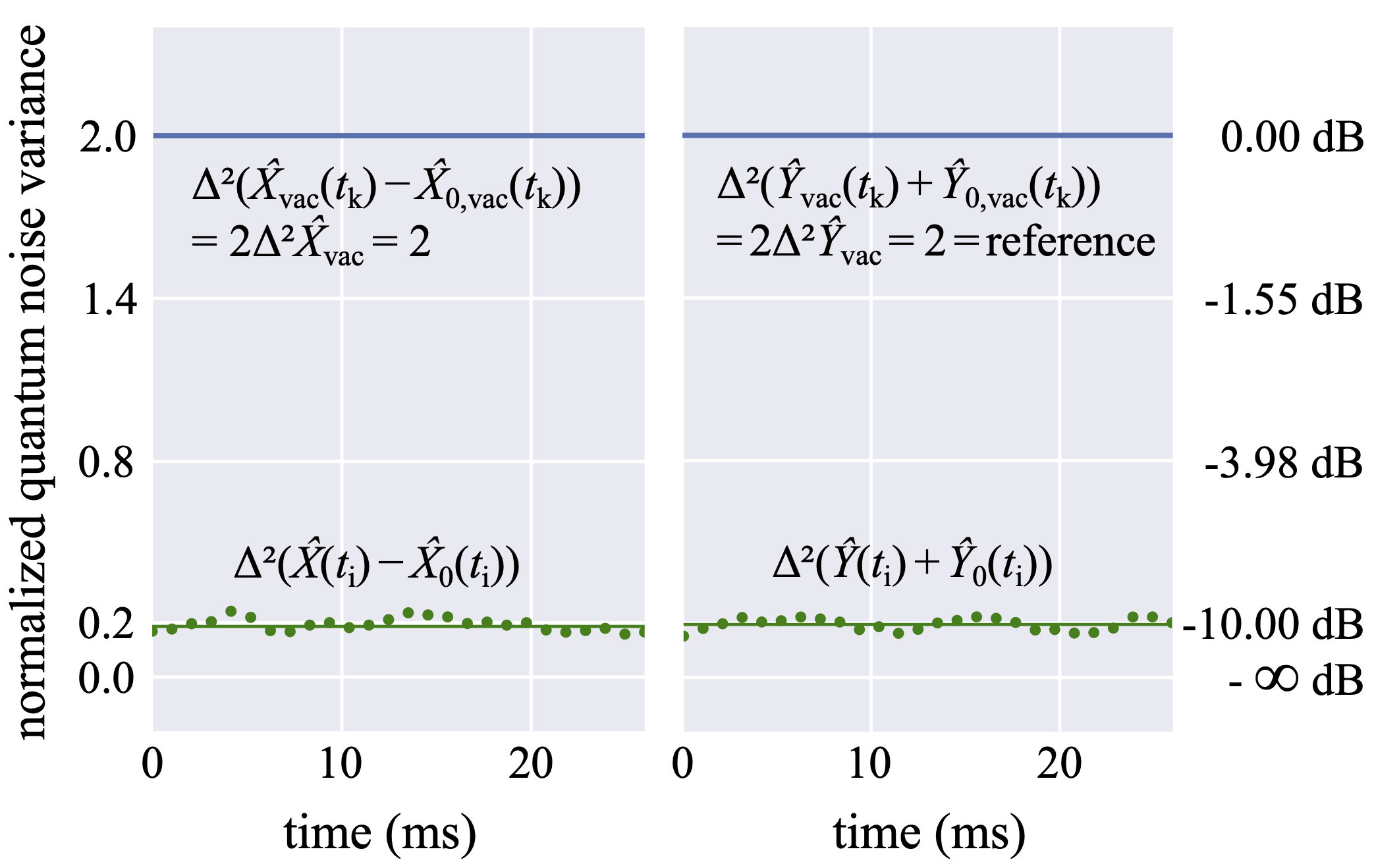}
         \vspace{-5mm}
    \caption{
Variances of $X(t_i)$-data (left) and $Y(t_i)$-data (right), which characterise the quality of our data sampling with respect to the entangled reference. Shown is the example of stationary zero displacement, i.e.~$\langle \hat X(t) \rangle = \langle \hat Y(t) \rangle= 0$. 
Solid lines correspond to the variance of 2600 measuring points each, without entanglement (top lines) and with entanglement (bottom lines). The latter corresponds to 10\,dB two-mode squeezing. The two types of modulations can be measured simultaneously with an uncertainty product of $\Delta (\hat X (t_i)) \Delta (\hat Y (t_i)) \approx 0.2$ violating Inequality\,(\ref{eq:hup2}) by a factor of $\sim\!10$. Here, $f = 5$\,MHz and $\Delta f = 20$\,kHz. Dots represent variances calculated over 260 consecutive measuring points. 
}
    \label{fig:3}
\end{figure}

\textit{Experimental setup} -- 
Fig.\,\ref{fig:2} shows the schematic of our experiment. A commercial erbium-doped fibre laser generated 1\,W of quasi-monochromatic light at the wavelength of 1550\,nm. About half of the light was frequency doubled to provide the pump light for two squeezed-light resonators. The latter used resonator-enhanced degenerate type\,0 optical-parametric amplification (OPA) in periodically poled potassium titanyl phosphate (PPKTP). The two output fields carried modulation spectra around 5\,MHz in squeezed vacuum states and were overlapped at balanced beam splitter BS$_{\rm 1}$. The result were two fields whose modulations were strongly EPR entangled, which was characterised in a precursor experiment \cite{Eberle2013}.
Here, we recombined the entangled beams on a second balanced splitter (BS$_3$). The optical path length difference was controlled to convert them back to two squeezed beams. Due to necessarily imperfect interference contrasts at the two beam splitters, the final squeeze factors could only be lower than the initial squeeze factors of the input modes (subscripts 1 and 2). 
The BHDs used optical local oscillators (LO's) of about 10\,mW from the joint fibre laser. The phase differences between the LO's and the squeezed fields were stably controlled to zero and ninety degrees, respectively. BHD$_1$ at zero degrees sampled values of a squeezed amplitude quadrature amplitude; BHD$_2$ at ninety degrees sampled values of a squeezed phase quadrature amplitude, both with a sampling frequency of 200\,MHz. 
To avoid aliasing, we applied an analogue lowpass-filter with a corner frequency of 50\,MHz to each channel. Post processing was done with a self-written Python script, which was used to digitally demodulate the data at $f = 5$\,MHz and subsequent finite impulse response (FIR)-lowpass-filtering with a cut off frequency of $\Delta f /2 = 10$\,kHz. 
Fig.\,\ref{fig:3} represents the entanglement quality of our setup in terms of variances. 

\begin{figure}[h!!!!!!!!!!!!!!!!!!!!!!!!!!!!!!!!!!!!!!]
     \vspace{0mm}
     \includegraphics[width=8.5cm]{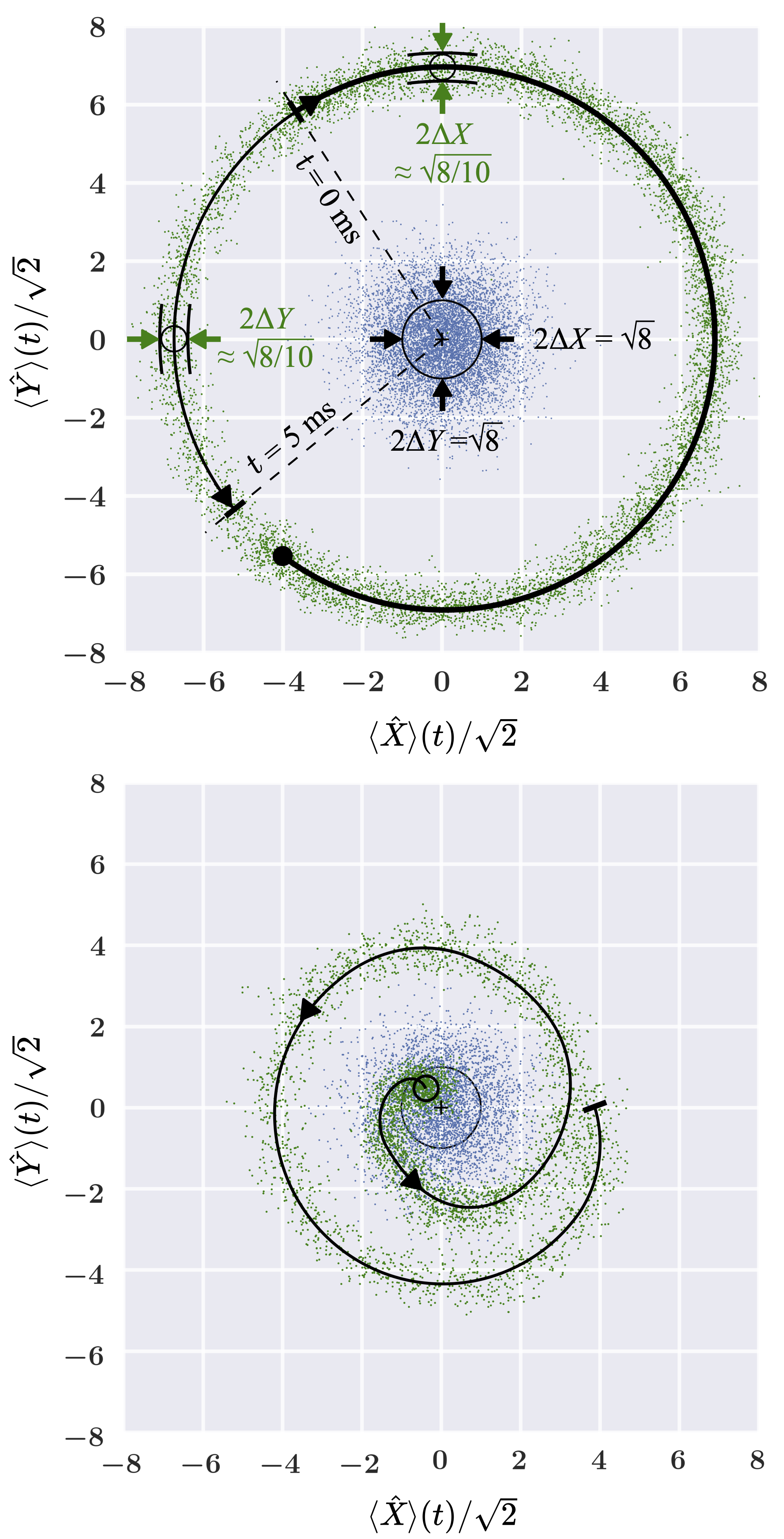}
     \vspace{-1mm}
    \caption{
Example phase-space trajectories of about 5\,ms length measured with sub-Heisenberg imprecision (solid line surrounded by green dots) in comparison to measurements on ground states (centred, blue dots). The dots represent single measurements ($(X\!-\!X_0)(t_i);(Y\!+\!Y_0)(t_i)$) performed at subsequent times $t_i$, with $t_{i+1} - t_i =10\mu$s. (To increase the number of points we superposed 15 and 8 identical measurements, respectively.) 
The spreads of the data points in the two phase space directions represent the relevant standard deviations of quantum noise in estimating the trajectories. The sub-Heisenberg uncertainty area is revealed by comparing the small circles to larger ones in the centres, which represent the lower bound in Inequality\,(\ref{eq:hup2}). The latter is surpassed by a factor of about ten. The upper trajectory represents a changing type of modulation at constant modulation depth. The bottom one additionally shows a continuously decreasing modulation depth. 
}
 \vspace{-7mm}
\label{fig:4}
\end{figure}

The time-dependent displacement $\alpha (t)$ in our setup corresponded to the time-dependent modulation at $f = 5$\,MHz of a coherent carrier field that was transmitted through the high reflectivity mirror BS$_{\rm 2}$ ($R$\,=\,99.99\%). The high reflectivity minimised decoherence, i.e.~optical loss to the entanglement. Changing the peak voltage to the electro-optical modulator (EOM, as shown in Fig.\,\ref{fig:2}), changed the absolute value of the displacement $|\alpha|$. Changing the DC voltage to the piezo-actuated phase-shifter ($U_1$) changed the differential excitation in $\langle \hat X \rangle$ and $\langle \hat Y \rangle$.
The time series produced at BHD$_1$ and BHD$_2$ represented simultaneous measurements of the system's conjugate displacement components with respect to the corresponding values of the (entangled) reference. 
Since $\langle \hat X_0 \rangle \!=\! \langle \hat Y_0 \rangle \!=\! 0$, the data serves for monitoring the trajectory $(\langle \hat X \rangle; \!\langle \hat Y \rangle)(t)$.

\textit{Experimental Results}.
Fig.\,\ref{fig:4} shows two phase-space trajectories $(\langle \hat X \rangle; \!\langle \hat Y \rangle) (t)$ (solid lines) measured with a precision suspending Inequalities\,(\ref{eq:hup1}) and (\ref{eq:hup2}).
Added are individual data points from simultaneous measurements of  $(X\!-\!X_0)(t_i)$ and $(Y\!+\!Y_0)(t_i)$, 
when the interrogated system was entangled with the reference system. 
Also shown are individual data points from simultaneous measurements of  $(X\!-\!X_0)(t_i)$ and $(Y\!+\!Y_0)(t_i)$, when the entanglement source was switched off and the modulations $\langle \hat X \rangle$ and $\langle \hat Y \rangle$ set to zero. These data points accumulated around the phase space origin and was used to derive the factor by which the Inequalities\,(\ref{eq:hup1}) and (\ref{eq:hup2}) were surpassed.
The standard deviations in $(X\!-\!X_0)(t_i)$ and $(Y\!+\!Y_0)(t_i)$ around the actual phase-space trajectories $(\langle \hat X \rangle; \!\langle \hat Y \rangle) (t)$ (solid line) were reduced by more than $\sqrt{10}$. This factor is highlighted by the different radii of the small circles.
The phase-space trajectories were thus tracked with an uncertainty product that violated Inequality\,(\ref{eq:hup2}) by slightly more than a factor of $10$.
As expected, the factor by which Heisenberg's uncertainty limit was surpassed directly corresponded to the strength of the entanglement. 
Increasing the entanglement strength requires further reduction of optical loss, including further increase of photo-electric detection efficiency \cite{Vahlbruch2016}.\\
Fig.\,\ref{fig:4}\,(a) represents a constant modulation depth, while the kind of modulation was continuously changed. The system had a pure amplitude modulation when $\langle \hat Y \rangle (t)=0 $ and a pure phase quadrature modulation when $\langle \hat X \rangle (t)=0 $. 
The amplitude of the AC voltage at the EOM ($U_2$) was constant and just the DC voltage at the piezo actuator ($U_1$) continuously changed. The trajectory started at about $(\langle \hat X \rangle; \!\langle \hat Y \rangle) = (- 3.7\sqrt{2}; 5.8\sqrt{2})$, completed almost a full cycle, returned, and stopped at about $(-5.3 \sqrt{2}; -4.3 \sqrt{2})$. 
The bottom panel shows another example trajectory whose modulation depth also changed, resulting in a phase and amplitude dependent trajectory.

\textit{Discussion and conclusion.} 
Our experiment demonstrates that any individual measurement of two non-commuting observables at the same time improved when entanglement is exploited. The observables considered are the real and imaginary parts of phase-space displacement, which describe the depths of the amplitude modulation and the phase quadrature around a selected radio frequency of a continuous-wave quasi-monochromatic laser beam.
The temporal change of the modulations depths corresponds to a trajectory in phase space. Our experiment shows that the trajectory can be monitored by a large number of individual measurement pairs with a tenfold reduced imprecision in each of the observables compared with the variances of measurements on the ground state (without entanglement). The reduction factors correspond to the squeeze factor in both observables (10\,dB) realised by the entanglement resource. 
We conclude that the time-varying displacement can in principle be monitored with arbitrary precision without averaging, i.e.~even for random walks.   
We thus further conclude that the often-quoted interpretation of Heisenberg's uncertainty relation `two non-commuting observables of a quantum system cannot be measured simultaneously with arbitrary precision' is incorrect. 
In light of our experiment, the statement becomes correct, if completed by `...with respect to a reference system that was or has been coupled to a thermal environment', since in this case the reference system cannot be quantum correlated. \\ 
The phase-space displacement in our experiment is overlapped with the entangled fields via a high reflectivity beam splitter. In principle the displacement 
can also be produced directly in the beam path of one of the entangled states by a combination of an amplitude and a phase modulator, because the entangled states are carried by accompanying monochromatic fields. An important issue is to keep the entanglement decoupled from the environment before the measurement.   
The reduction factor in the quantum imprecision achieved is of practical significance and support the emergent field of quantum sensing. In gravitational-wave observatories, entangled light provides additional sensitivity improvements compared to squeezed light \cite{LSC2011,Tse2019, Acernese2019,Schnabel2010} by mitigating disturbances from back-scattered light \cite{SteinlechnerS2013} and from quantum radiation pressure \cite{Ma2017,Suedbeck2020,Yap2020}. The setup realised constitutes state of the art quantum optics technology suitable for the generation and detection of Gaussian cluster states for measurement-based quantum computing \cite{Raussendorf2001,OBrian2007,Larsen2019}.\\ [4mm]
{\bf Acknowledgment}\\
The authors thank Mikhail Korobko for useful comments on the manuscript.\\[3mm]
{\bf Author contributions}\\
J.Z. performed the experiments, analysed the data, and wrote the manuscript. 
R.S. conceived the experiment, supported data analysis, and wrote the manuscript.\\[3mm]
{\bf Competing interests}\\
The authors declare that there are no competing interests.\\[3mm]
{\bf Additional information}\\
Supplementary information about quadrature field observables.\\
{\bf Data availability statement}\\
The data sets generated during the current study are available from the corresponding author on reasonable request.\\

\end{document}